\documentclass[%
 reprint,
superscriptaddress,
 amsmath,amssymb,
 aps,
]{revtex4-2}

\usepackage{graphicx}% Include figure files
\usepackage{dcolumn}% Align table columns on decimal point

\usepackage{bm}% bold math
\usepackage{braket}
\usepackage{xcolor}
\begin{document}

\preprint{APS/123-QED}

\title{Optical microcavities as platforms for entangled photon spectroscopy}% 
%\thanks{A footnote to the article title}%

\author{Ravyn Malatesta}
\affiliation{%
School of Chemistry and Biochemistry and School of Physics, Georgia Institute of Technology, Atlanta, Georgia 30332, USA.}%

\author{Lorenzo Uboldi}
\affiliation{%
 Department of Physics and Center for Functional Materials,  Wake Forest University, 2090 Eure Drive, Winston Salem, North Carolina 27109, USA. 
}%
\affiliation{Dipartimento di Fisica, Politecnico di Milano, Piazza Leonardo da Vinci 32, Milano, Italy.}%

\author{Evan J. Kumar}
\affiliation{%
 Department of Physics and Center for Functional Materials,  Wake Forest University, 2090 Eure Drive, Winston Salem, North Carolina 27109, USA.   
}%

\author{Esteban Rojas-Gatjens}
\affiliation{%
School of Chemistry and Biochemistry and School of Physics, Georgia Institute of Technology, Atlanta, Georgia 30332, USA.}
\affiliation{%
 Department of Physics and Center for Functional Materials,  Wake Forest University, 2090 Eure Drive, Winston Salem, North Carolina 27109, USA. 
}%

\author{Luca Moretti}
\affiliation{Dipartimento di Fisica, Politecnico di Milano, Piazza Leonardo da Vinci 32, Milano, Italy.}%

\author{Andy Cruz}
\affiliation{Department of Physics, City College of New York, City University of New York, NY 10031, USA.}
\author{Vinod Menon}
\affiliation{Department of Physics, City College of New York, City University of New York, NY 10031, USA.}

\author{Giulio Cerullo}
\affiliation{Dipartimento di Fisica, Politecnico di Milano, Piazza Leonardo da Vinci 32, Milano, Italy.}%
%\affiliation{
% Department of Physics and Center for Functional Materials, Wake Forest University
%}%
\author{Ajay~Ram~Srimath~Kandada}
\email[Electronic address: ]{srimatar@wfu.edu}
\affiliation{%
 Department of Physics and Center for Functional Materials,  Wake Forest University, 2090 Eure Drive, Winston Salem, North Carolina 27109, USA. 
}%

\date{\today}

\begin{abstract}
Optical microcavities are often proposed as platforms for spectroscopy in the single- and few-photon regime due to strong light-matter coupling. For classical-light spectroscopies, an empty microcavity simply acts as an optical filter. However, we find that in the single- or few-photon regime treating the empty microcavity as an optical filter does not capture the full effect on the quantum state of the transmitted photons. Focusing on the case of entangled photon-pair spectroscopy, we consider how the propagation of one photon through an optical microcavity changes the joint spectrum of a frequency-entangled photon pair. Using the input-output treatment of a Dicke model, we find that propagation through a strongly coupled microcavity above a certain coupling threshold enhances the entanglement entropy between the signal and idler photons. These results show that optical microcavities are not neutral platforms for quantum-light spectroscopies and their effects must be carefully considered when using change in entanglement entropy as an observable.

\end{abstract}

%\keywords{Suggested keywords}%Use showkeys class option if keyword
                              %display desired
\maketitle

%\tableofcontents

\section{\label{sec:intro}Introduction}
Due to spectacular advances within the field of quantum optics, experimentalists can now control non-classical states of light with high levels of precision in optics laboratories. Combined with advances in single-photon detection, these innovations lay the groundwork for the growing field of quantum-light spectroscopy\cite{roadmap2020}. There are many advantages to using quantum-light for spectroscopy, including access to information otherwise inaccessible using classical spectroscopies\cite{li2017}\cite{cuevas2018}\cite{asban2021}, and importantly superior signal-to-noise ratio that can enable spectroscopy at extremely low excitation fluence. \\

Quantum light refers to any state of light that cannot be described classically, such as single photons, squeezed light, or entangled photon pairs. Entangled photon pairs exhibit non-classical correlations that provide an advantage for both linear and nonlinear spectroscopies\cite{gea1989}\cite{oka2010}\cite{szoke2020}\cite{schlawin2013}\cite{dorfman2016}. In the single- or few-photon regime, classical spectroscopic signals are swamped with noise but entanglement-enhanced spectroscopies can surpass the shot-noise limit by taking advantage of quantum correlations\cite{dorfman2021}\cite{kandada2021}. Similarly, entangled light can enhance signal-to-noise ratios of nonlinear spectroscopies, resulting in sharper spectroscopic features and greater simultaneous time-frequency resolution\cite{raymer2013}. Furthermore, entangled-photon pairs provide direct access to nonlinear processes even at low-level excitation, facilitating the study of nonlinear processes in photo-sensitive systems that might bleach or otherwise be destroyed at excitation powers\cite{kojima2004}. Hao Li \textit{et al.} describe theoretically how the entanglement entropy of biphoton states (photon pair states) can be used as a probe of many-body correlations that are often elusive or obscured in classical nonlinear spectroscopic measurements\cite{li2019}.\\

In the single- or few-photon regime, a challenge arises for spectroscopists because of the low proability of light-matter interactions. One popular method to address this problem is to use an optical microcavity to couple to optical excitations in materials and thus enhance the processes of interest\cite{carnio2021}\cite{levinsen2019}\cite{osaka2014}\cite{gu2020}. Optical microcavities are extremely controllable platforms for light-matter interaction; they are used to manipulate molecular states, enhance spontaneous emission, and drive chemical reactions\cite{flick2017}. For all of their uses, microcavities are an extremely versatile platform for quantum spectroscopy, but they are not neutral platforms and cannot be treated as such.\\

To demonstrate this, we consider how the joint spectrum and entanglement entropy of a frequency-entangled biphoton state changes after one photon (the idler) propagates through an optical microcavity. We first briefly describe the modeling of the biphoton joint spectrum and its transformation using input-output theory. We then consider an empty microcavity. Although for classical light an empty microcavity behaves as a simple optical filter, we find experimentally that treating the microcavity as an optical filter does not capture the full effect on the transmitted biphoton state. With simple input-output theory, we can model the filtering effect of the empty microcavity but cannot explain the full joint spectral transformation. We next move on to a simple model of a microcavity coupled to $N$ two-level systems and consider how strong-coupling transforms the joint spectrum and entanglement entropy of the biphoton state after the idler passes through an active microcavity. We find that above a certain coupling strength, passing through the microcavity system alone, regardless of detuning, enhances the entanglement entropy even without including many-body interactions in the model. These results confirm that optical microcavities are not neutral platforms for quantum-light spectroscopies and their effects must be carefully considered when using change in entanglement entropy as a spectroscopic observable.

\section{\label{sec:biphoton}Biphoton state transformation}
\begin{figure}
    \includegraphics[scale=0.37]{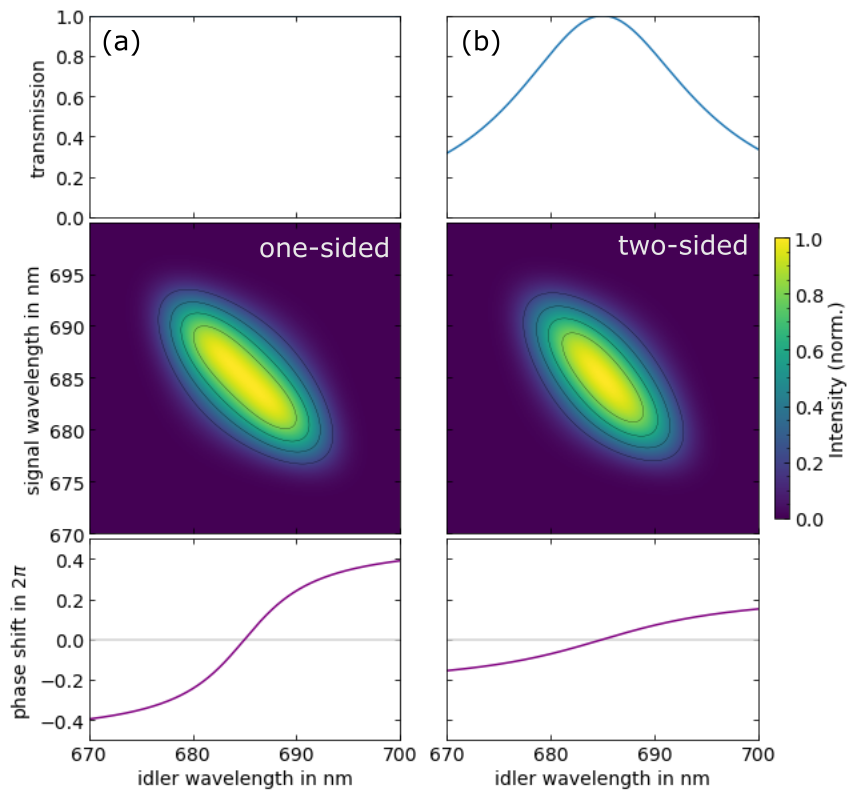}
    \caption{Transmission function, joint spectral intensity, and applied phase shift of a (a) one-sided and (b) two-sided empty microcavity following Gardiner and Collett.}
    \label{fig:CnG}
\end{figure}

\subsection{\label{SPDC}Joint spectrum and entanglement}
Sources of entangled photons that are based on spontaneous parametric downconversion (SPDC) generate two daughter photons, historically called the \textit{signal} and \textit{idler}, from a single pump photon according to energy- and momentum-conservation. Following the development of Zielnicki \textit{et al.}\cite{zielnicki2018}, a generic biphoton state of a signal and idler pair can be written as
%------------------------------------------------------------------------------------
\begin{eqnarray}
\label{ket_initial}
\ket{\psi_{s,i}} = \int\int d\omega_s d\omega_i \mathcal{F}\left(\omega_s, \omega_i \right)a_1^{\dag}(\omega_s)a_2^{\dag}(\omega_i)\ket{0},
\end{eqnarray}
%------------------------------------------------------------------------------------
where the creation operators $a_1^{\dag}(\omega_s)$ and $a_2^{\dag}(\omega_i)$ operate on the vacuum state to create photons at frequency $\omega_s$ and $\omega_i$, respectively. The joint spectral amplitude, $\mathcal{F}\left(\omega_s, \omega_i \right)$, describes the frequency-correlations between the signal and idler photons.\\

Experimentally, we typically measure the joint spectral intensity (JSI),
%------------------------------------------------------------------------------------
\begin{eqnarray}
\label{jsi}
|\mathcal{F}\left(\omega_s, \omega_i \right)|^2 = |A\left(\omega_s, \omega_i\right)|^2|\Phi\left(\omega_s, \omega_i\right)|^2,
\end{eqnarray}
%------------------------------------------------------------------------------------
where $A\left(\omega_s, \omega_i\right)$ is based on the spectral amplitude of the pump beam and $\Phi\left(\omega_s, \omega_i\right)$ is determined by the \textit{phase-matching conditions} and \textit{spatial profile} of the pump.\\ 

To quantify the entanglement between the signal and idler photons, we compute the von Neumann entanglement entropy, $S$. We first normalize the joint spectral amplitude, and then use singular value decomposition to find the Schmidt coefficients $\lambda_j$ which satisfy the normalization condition $\sum_j \lambda_j^2 = 1$. We then calculate the entanglement entropy as
%------------------------------------------------------------------------------------
\begin{eqnarray}
\label{entropy}
S = - \sum_i \lambda_i^2\ln\left(\lambda_i^2\right).
\end{eqnarray}
%------------------------------------------------------------------------------------

\subsection{\label{inout}Application of input-output theory}
In their seminal work in 1984~\cite{collett1984}, Collett and Gardiner develop a general input-ouput theory that relates output operators to input operators via internal dynamics of a cavity system governed by quantum Langevin equations. For a single photon mode, if we express the input-output transformation as a frequency-dependent function $C(\omega_i)$, we simply write the output creation operator in terms of the input as
%------------------------------------------------------------------------------------
\begin{eqnarray}
\label{output_operator}
\tilde{a}^{\dag}(\omega_i)=C\left(\omega_i\right)a^{\dag}(\omega_i).
\end{eqnarray}
%------------------------------------------------------------------------------------\\

Now, considering the case of a biphoton state where we allow only the idler photon to propagate through a microcavity system, we replace the original idler creation operator $a_2^{\dag}(\omega_i)\rightarrow\tilde{a}_2^{\dag}(\omega_i)$ to get the transformed biphoton state
%------------------------------------------------------------------------------------
\begin{eqnarray}
\label{ket_final}
\ket{\Psi} = \int\int d\omega_s d\omega_i \mathcal{F}\left(\omega_s, \omega_i \right)C\left(\omega_i\right) a_1^{\dag}(\omega_s)a_2^{\dag}(\omega_i)\ket{0}.
\end{eqnarray}
%------------------------------------------------------------------------------------
In this simple approach, the transformed JSA is the product $\mathcal{F}\left(\omega_s, \omega_i \right)\cdot C\left(\omega_i\right)$. The transformed state bears similarity to the expression developed by Kalashnikov \textit{et al.} before they act with a beamsplitter to see how interaction with a resonant medium changes the quantum interference pattern in Hong-Ou-Mandel interferometry\cite{kalashnikov2017}.

\section{\label{sec:empty}Propagation through an empty microcavity}
\subsection{\label{sec:emptytheory}Theory}
Following the input-output formalism of Collett and Gardiner\cite{collett1984}, we consider an empty cavity confining a single optical mode. Starting with a one-sided empty microcavity, i.e. a microcavity with substantial loss through a single mirror, the transformation of the output photon creation operator in terms of the input is
%------------------------------------------------------------------------------------
\begin{eqnarray}
\label{1sided}
\tilde{a}^\dag\left(\omega\right) = \frac{\frac{1}{2}\gamma-i\left(\omega-\omega_0\right)}{\frac{1}{2}\gamma+i\left(\omega-\omega_0\right)}a^\dag\left(\omega\right),
\end{eqnarray}
%------------------------------------------------------------------------------------
where $\gamma$ is the coupling strength of the cavity photons to input(output) photons and $\omega_0$ is the frequency of the cavity mode. The coupling strength $\gamma$ is directly related to the cavity photon lifetime, $\tau = 1/\gamma$. For all our simulations, we choose $\gamma$ such that the cavity photon lifetime $\tau$ is 150\,fs.\\

As noted by Collett and Gardiner, the one-sided cavity imposes a frequency-dependent relative phase shift, but does not change the JSI. Therefore the JSI shown in Fig. \ref{fig:CnG}(a) is identical to that of the input biphoton state. For all simulations shown here, we use the same input state, assuming a Gaussian pump with a central down-converted wavelength of 685\,nm for both signal and idler photons. To replicate experimental conditions, we apply detection filters to both the signal and idler. For the filter shape, we choose a Gaussian squared, centered at 685\,nm with an 8\,nm bandwidth. Until we consider the effect of the pump bandwidth on entanglement entropy, the pump bandwidth is set at 6\,nm.\\

Next, we move on to a two-sided empty microcavity, i.e. a cavity with leaky mirrors on both sides, and so with two input and two output modes. We assume the coupling to be same for both mirrors, $\gamma_1 = \gamma_2 = \gamma$, and a single input mode. Thus in transmission, the output photon creation operator is
%------------------------------------------------------------------------------------
\begin{eqnarray}
\label{2sided}
\tilde{a}^\dag\left(\omega\right) = \frac{\gamma}{\gamma+i\left(\omega-\omega_0\right)}a^\dag\left(\omega\right).
\end{eqnarray}
%------------------------------------------------------------------------------------
Now we see the filtering effect of the empty microcavity acting on the idler photon, as shown in Fig. \ref{fig:CnG}(b). The centering and bandwidth of the transmission function are determined by the cavity mode frequency and the cavity photon lifetime, respectively. The filtering effect of the microcavity slightly affects the entanglement entropy bringing the base entanglement entropy from $S=0.395$ to $S=0.359$. 
\begin{figure}
    \includegraphics[scale=0.37]{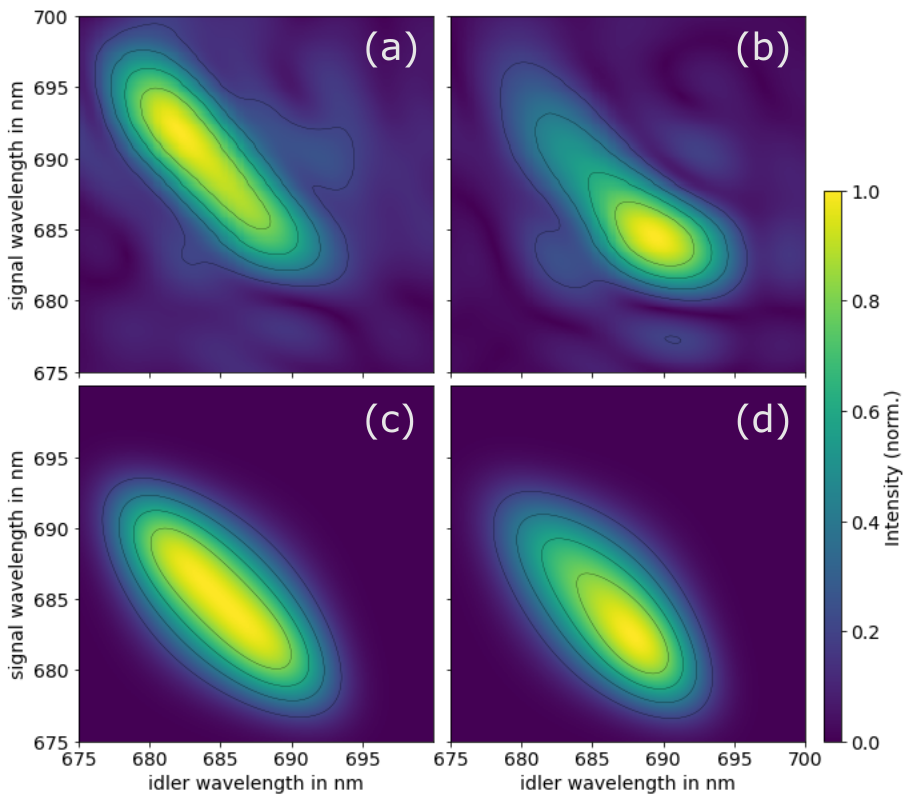}
    \caption{Joint spectral intensity before and after propagating through an empty microcavity from (a, b) experimental measurement and (c, d) input-output theory.}
    \label{fig:expt}
\end{figure}

\subsection{\label{emptyexpt}Experiment}

To further test the formalism developed in the previous section, we experimentally measure the JSI of a biphoton state with one of the photons transmitted through an empty microcavity. The spectrally entangled state is generated in a Type-I $\beta$-Barium Borate (BBO) crystal phase-matched for SPDC close to the degeneracy at the pump wavelength of 343\,nm. The pump beam here is the third harmonic of a femtosecond laser oscillator (Pharos, Light Conversion) output operating at 1030\,nm and 75\,MHz. The photons are spatially separated and transmitted through a translating-wedge-based identical pulses encoding system (GEMINI, Nireos srl) and a co-incidence detection system (Hydraharp, Picoquant), which enable measurement of spectral correlations between the photons. More details of the measurement system can be found in Ref.~\citenum{moretti2023}. The JSI spectrum of the as-prepared biphoton state is shown in Fig. \ref{fig:expt}(a), in which the spectral correlation between the signal and idler photons is evident through the diagonal feature.

We transmit the idler photon of this state through a planar optical microcavity, which is built on distributed Bragg reflectors (DBR) and has an optical resonance at 691\,nm with a full width at half maximum of 8\,nm at normal incidence. The JSI spectrum of the transmitted biphoton state is shown in Fig.~\ref{fig:expt}(b). We observe clear spectral filtering of the biphoton state with the peak of the JSI map at the peak resonance of the cavity. On closer inspection, we observe a reduction in the degree of spectral correlation in the transmitted state with the previously extended diagonal feature flattening along the idler-axis, close to the cavity resonance. To reproduce this behavior we consider a biphoton state whose JSI spectrum follows Eq.~\ref{jsi}, and shown in Fig.~\ref{fig:expt}(c). Based on the formalism developed in the previous section, we estimate the JSI spectrum of the biphoton state whose idler photon is transmitted through a microcavity. By setting the cavity resonance to 690\,nm in our simulation, we can approximate the experimentally measured transmission function of the empty microcavity. While the filtering effect is reproduced, we miss the effects of the microcavity that depend on the joint spectrum - that is we miss the effects of the microcavity that depend simultaneously on the signal and idler frequency even though only the idler propagated through the microcavity.

%\textcolor{red}{Here is where the experimental measurement description and discussion will go.}\\
\begin{figure*}
    \includegraphics[scale=0.5]{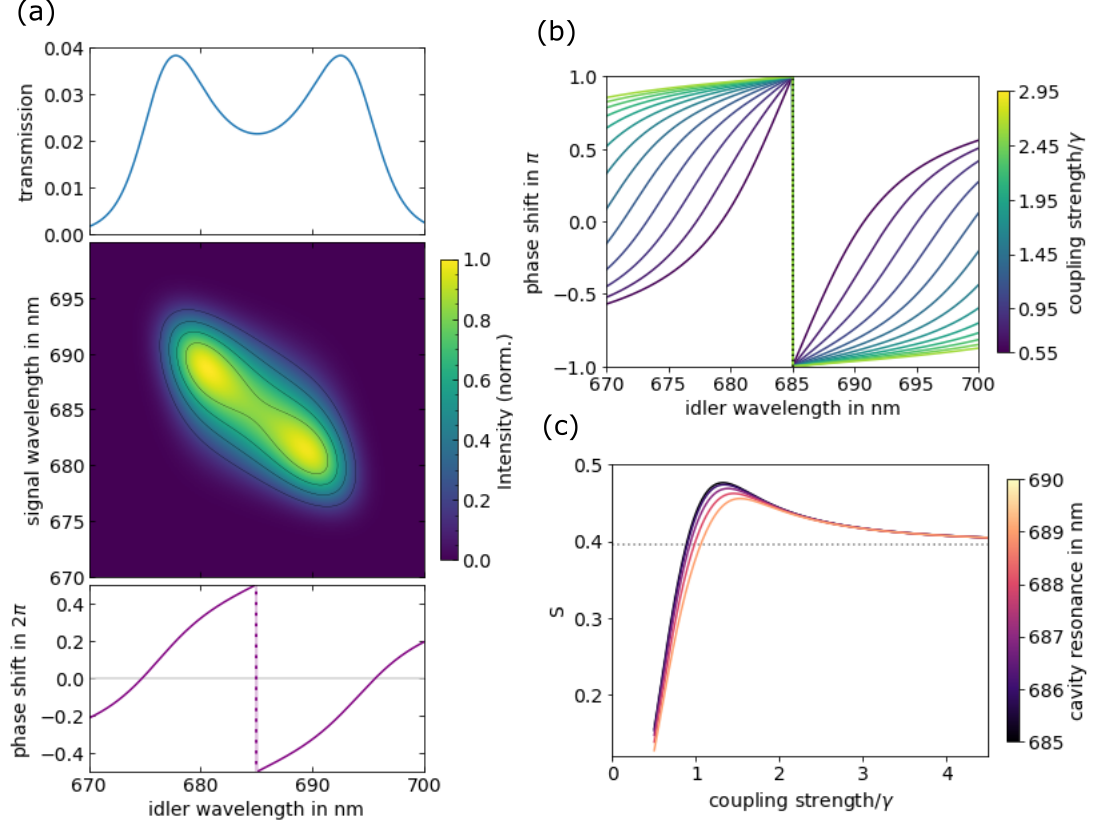}
    \caption{Changes to the biphoton state due to idler propagation through a strongly-coupled microcavity. (a) Transmission function, joint spectral intensity, and applied phase shift of a strongly-coupled ($\lambda=\gamma$) microcavity with a 150 fs lifetime with zero detuning, (b) dependence of the microcavity induced phase shift on coupling strength, and (c) change in entanglement entropy with coupling strength for variable cavity detuning for a molecular resonance at 685 nm, compared to input state entropy (dotted gray).}
    \label{fig:dicke_model}
\end{figure*}

\section{\label{sec:dicke}Propagation through a strongly-coupled microcavity}
Having already established that an empty microcavity has a non-trivial effect on the entanglement of frequency-entangled photon pairs, we now consider a simple model of a strongly-coupled microcavity system. Taking inspiration from Li \textit{et al.}\cite{li2017}, we use a Dicke model of $N$ identical 2-level emitters coupled to an optical cavity described by the following Hamiltonian:
%------------------------------------------------------------------------------------
\begin{eqnarray}
\label{ham}
\hat{H} &=& \sum_j \frac{\hbar\omega_e}{2}\hat{\sigma}_{z,j} + \sum_k \hbar(\omega_k - i\gamma)\hat{\psi}^\dag_k\hat{\psi}_k \nonumber \\
&& +\sum_{k,j} \frac{\hbar\lambda_{kj}}{\sqrt{N}}(\hat{\psi}^\dag_k + \hat{\psi}_k)(\hat{\sigma}^+_j + \hat{\sigma}^-_j),
\end{eqnarray}
%------------------------------------------------------------------------------------
where $\omega_e$ is the frequency of the emitter, $\{\hat{\sigma}_{z,j}, \hat{\sigma}_j^\pm\}$ are the corresponding spin-1/2 operators for site $j$, $\hat{\psi}^\dag_k$ is the cavity photon creation operator, and $\lambda_{kj}$ is the coupling between a cavity photon and a molecular excitation at site $j$. As before, $\gamma$ is the coupling of the cavity photon mode to an external photon mode. For simplicity, we consider only the normal cavity mode, $k=0$, with frequency $\omega_0$. We also constrain ourselves to the strong coupling regime, $\lambda > \gamma/2$, but stay well below the critical point $\lambda_c$ where the system undergoes a quantum phase transition.\\

Using an input-output treatment of this model, Li \textit{et al.} develop an analytical expression for the response function of the strongly-coupled system \cite{li2017} which we use to define the transformation function $C(\omega_i)$ for an idler photon propagating through the strongly-coupled microcavity. The transformation function depends on many parameters: the frequency of the molecular emitter $\omega_e$, the frequency of the cavity mode $\omega_0$, the cavity lifetime $1/\gamma$, and the strength of the coupling $\lambda$. \\

Applying this transformation to the same input biphoton state as before (Fig.~\ref{fig:expt}(c)), we immediately find much different behavior than for transmission through an empty microcavity. We analyze transmission through a strongly-coupled microcavity with both the molecular and cavity resonance at 685 nm, a 150 fs cavity photon lifetime, and equal coupling to the molecular excitation and external photons ($\gamma=\lambda$). The resulting transmission spectrum and the JSI map shown in Fig. \ref{fig:dicke_model}(a) are composed of two peaks associated with the lower and upper polariton states. While this is an expected result, two intriguing details emerge on deeper analysis. Firstly, we see sharp discontinuity in the applied phase shift corresponding to the molecular resonance. % has which shows clearly that The resulting biphoton state now has an entanglement entropy of $S=0.437$, a higher value than for the entanglement entropy of the input state. Upper and lower polariton peaks appear in the transmission function and in the joint spectral intensity, along with a sharp discontinuity in the applied phase shift corresponding to the molecular resonance. 
As the strength of the light-matter coupling increases with respect to the coupling to external photons, the applied phase shift begins to resemble a step function, as shown in Fig. \ref{fig:dicke_model}(b).\\

%\subsection{\label{sec:}Enhancement of entanglement entropy}
Secondly, the entanglement entropy of the transformed state is $S=0.437$, which is a higher value than the entanglement entropy of the input state $S=0.395$. But the increase in the entropy is curiously not monotonically related to the strength of light-matter coupling. %Curiously, 
The transformation of the biphoton state due to propagation of the idler photon through the strongly-coupled microcavity system \textit{increases} only above a certain coupling-strength threshold, see Fig. \ref{fig:dicke_model}(c). Below this threshold, the entropy substantially reduces, possibly due to the spectral filtering of the idler photons by the dominant molecular transition.  %  the entanglement entropy of the signal and idler pair. 
Of course, the exact coupling strength at which the enhanced entanglement entropy surpasses that of the input depends on the specific state and microcavity system parameters including the molecular resonance, cavity photon lifetime, and cavity detuning. In general, across several cavity detunings, shown in Fig. \ref{fig:dicke_model}(c), at the lower end of the strong-coupling limit when coupling to external photons out-competes coupling to the molecular excitation ($\gamma>\lambda>\gamma/2$), propagation through the coupled microcavity system suppresses the entanglement entropy even below propagation through an empty microcavity. %As shown in Fig. \ref{fig:entropy}, 
As the coupling strength increases, we reach a regime where the microcavity system improves the entanglement entropy past that of the input state for a wide range of cavity-detuning values, until the dependence of the entanglement entropy on cavity-detuning plateaus.\\
\begin{figure}
    \includegraphics[scale=.5]{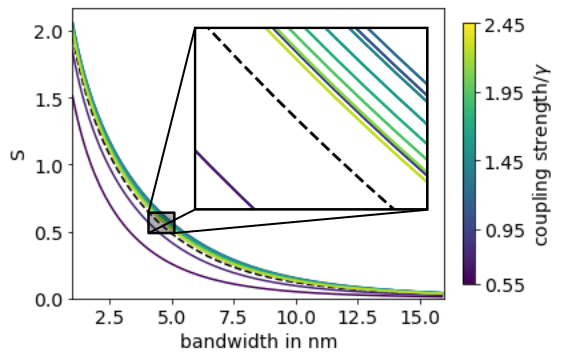}
    \caption{Entanglement entropy by pump bandwidth for increasing coupling strength compared to an empty microcavity (dashed black).}
    \label{fig:inset}
\end{figure}

We find a similar ebb and flow of entropy improvement when considering how the entanglement entropy changes with the bandwidth of the Gaussian pump generating the input biphoton state, seen in Fig. \ref{fig:inset}. For very narrow pump bandwidths, the frequencies of the signal and idler are strongly anti-correlated and the input state thus has a relatively high entanglement entropy. Within this limit, propagation of the idler through a sufficiently strongly coupled microcavity system still improves the entanglement entropy. Beyond $\lambda=1.35\gamma$, the benefit weakens but the entanglement entropy remains firmly above that for an empty microcavity.

\section{Conclusion}
In summary, we show with experiment and simple input-output theory that propagation through empty optical microcavities exerts a non-trivial effect on the state of frequency-entangled biphotons. We also theoretically consider the case of strongly coupled microcavities and identify peculiar transformations of the spectral correlations of the output biphoton state. From our experimental measurements of an empty microcavity we expect there to be further correlated effects in these systems that our simple input-output approach does not capture, but understanding the interplay of the modeled and unmodeled changes requires further theoretical development and experimentation. Nevertheless, we show even with a simple theoretical treatment that the microcavity platform has notable effects on entanglement entropy of biphoton states.\\

Notably, the experimental configuration and the model we consider here simply correspond to the \textit{linear} response of microcavities. Previous works propose to use the entanglement entropy of the biphoton state transmitted through the cavity as a probe of many-body processes, including polariton-polariton interactions. While these treatments show the biphoton entanglement entropy is sensitive to such many-body interactions, we have to also consider the non-trivial entropy changes identified in this work that can manifest even in the absence of any correlating mechanisms. While optical microcavities can indeed be excellent platforms that enable spectroscopy with entangled photons, care must be taken to design systems with light-matter coupling strengths that minimize the linear-response induced variations in the JSI, so that the transformation of the biphoton state can be directly correlated with many-body dynamics. \\

%\begin{figure}
%    \includegraphics[scale=.5]{fig4.png}
%    \caption{Entanglement entropy by cavity detuning for a molecular excitation at centered 685 nm as coupling strength increases compared to an empty microcavity (dotted gray).}
%    \label{fig:entropy}
%\end{figure}

\begin{acknowledgments}
A.R.S.K. acknowledges the start-up funds provided by Wake Forest University and funding from the Center for Functional Materials and the Office of Research and Sponsored Programs at WFU. The authors thank Prof Carlos Silva, Prof Eric Bittner and Dr Andrei Piryatinski for insightful discussions. This material is based upon work supported by the National Science Foundation Graduate
Research Fellowship under Grant No. DGE-2039655. Any opinion, findings, and conclusions or recommendations expressed in this material are those of the authorss and do not necessarily reflect the views of the National Science Foundation.
\end{acknowledgments}

%\bibliography{EPPcavity}% Produces the bibliography via BibTeX.
%merlin.mbs apsrev4-1.bst 2010-07-25 4.21a (PWD, AO, DPC) hacked
%Control: key (0)
%Control: author (72) initials jnrlst
%Control: editor formatted (1) identically to author
%Control: production of article title (-1) disabled
%Control: page (0) single
%Control: year (1) truncated
%Control: production of eprint (0) enabled
%
\end{document}